# An Implementation on Detection of Trusted service provider in Mobile Ad-hoc Networks

Mr.Rahul A Jichkar[#1], Dr.M.B. Chandak[*2]

[#]*M.Tech Scholar, Department of Computer Science,*
*Shri Ramdeobaba College of Engg. & Management, Nagpur, India.*
[*]*Head, Department of Computer Science,*
*Shri Ramdeobaba College of Engg. & Management, Nagpur, India.*

*Abstract*— The mobile ad-hoc network consists of energy constraint devices called nodes communicating through radio signals forming a temporary network i.e. nodes are continuously switching from one network to another. To minimize the power consumption, we form the clusters with an elected cluster head (Service Provider) based on any cluster head selection strategy. As the topology of this network is time variant attributable to node mobility, nodes are continuously leaving and entering the clusters, automatically registering with the cluster head to become the member of the cluster. But, there can be a scenario where a new node wants to access a service provided by the cluster head, at this time the newly entered node is unaware of the trustworthiness of the cluster head. To establish a trusted link amongst newly entered node and CH we have adopted an indirect trust computation technique based on recommendations, which form an important component in trust-based access control models for pervasive environment. It can provide the new node the confidence to interact with unknown service provider or CH to establish a trusted link for reliable accessibility of the service. In this paper, we shall present some existing indirect trust based techniques and subsequently discuss our proposal along with its merits and future scope.

## I. INTRODUCTION

Mobile Ad Hoc Networks (MANETs) are composed of fully autonomous wireless devices forming a temporary networks, these are suitable in the environment where the infrastructure is not fixed or we can say that infrastructure is time variant. In case of fixed hard-wired networks attacks are predictable with physical defense at firewalls and gateways, whereas, attacks on MANETs can come from any directions and may target any node. Due to dynamic topology of the networks any security solution with static configuration are not sufficient. Any node participating the network should not be directly trusted without verifying its trust information. If the trust information is available for each and every node in the network, then it is convenient to take precautionary measures to prevent the attacks using appropriate intrusion detection techniques. Moreover, it will be more sensible to reject or ignore hostile service requests. As the overall environment in MANET is cooperative by default, these trust relationships are extremely susceptible to attacks. So, in order to avoid the overhead of handling the network as a whole, nodes are grouped into clusters.

There are several cluster formation strategies which are used to form the clusters, most of these techniques are based on the degree of connectivity of a particular node. A node having highest degree of connectivity i.e. it is surrounded by maximum neighbours, then the node is elected as cluster head. There are also some other parameters such as, energy, node-id, etc. which can be taken into consideration while electing a cluster head. For simplicity we will be considering the static mobility model with heterogeneous nodes i.e. each node will be having different capabilities. The node having more computing power and battery life can be elected as a cluster head, as it will handle inter cluster communication and computation.

## II. RELATED WORK

The dynamism of pervasive computing environment allows ad hoc interaction of known and unknown autonomous entities that are unfamiliar and possibly hostile. In such environment where the service requesters have no personal experience with unknown service providers (here cluster heads), trust and recommendation models are used to evaluate the trustworthiness of unfamiliar entities. Recently, research in designing defense mechanisms to detect dishonest recommendation in these open distributed environments has been carried out [1-18]. The defense mechanisms against dishonest recommendations has been grouped into two broad categories, namely exogenous method and endogenous method [1].The approaches that fall under endogenous method use other external factors along with the recommendations (reputation of recommender and credibility of recommender) to decide the trustworthiness of the given recommendation. However, these approaches assume that only highly reputed recommenders can give honest recommendations and vice versa. In endogenous method, the recommendation seeker has no personal experience with the entity in question. It relies only on the recommendations provided by the recommender to detect dishonest recommendation. The method believes that dishonest recommendations have different statistical patterns from honest recommendations. Therefore, in this method, filtering of dishonest recommendation is based on analyzing and comparing the recommendations themselves. In trust models where indirect trust based on recommendations is used only once to allow a stranger entity to interact, endogenous method based on the majority rule is commonly used. Dellarocas [13] has proposed an approach based on controlled anonymity to separate unfairly high ratings and fair ratings. This approach is unable to handle unfairly low ratings [14]. In [15], a filtering algorithm based on the beta distribution is proposed to determine whether each recommendation R$i$ falls between q quartile (lower) and (1 − q) quartile (upper). Whenever a recommendation does not lie between the lower and upper quartile, it is considered malicious and its recommendation is excluded. The technique assumes that recommendations follow beta distribution and is effective only if there are effectively many recommendations. Weng et al. in [16] proposed a filtering mechanism based on entropy. The basic idea is that if a recommendation is too different from majority opinion, then it could be unfair. The approach is similar to other reputation-based models except that it uses entropy to differentiate between different recommendations. A context-specific and reputation-based trust model for pervasive computing environment was proposed [17] to detect malicious recommendation based on control chart method. The control chart method uses mean and standard deviation to calculate the lower confidence limit (LCL) and upper confidence limit (UCL). It is assumed that the recommendation values that lie outside the interval defined by LCL and UCL are malicious, therefore discarded from the set of valid recommendations. It considers that a metrical distance exists between valid and invalid recommendations. As a result, the rate of filtering out the false positive and false negative recommendation is really high. Deno et al. [18] proposed an iterative filtering method for the process of detecting malicious recommendations. In this model [18], an average trust value ($T_{avg}$) of all the recommendations received (TR) is calculated. The inequality | $T_{avg}$(B) − TR(B) | > S, where B is the entity for which recommendations are collected from i recommenders (R) and S is a predefined threshold in the interval [0 1], is evaluated. If that inequality holds, then the recommendation is false and is filtered out. The method is repeated until all false recommendations are filtered out. The effectiveness of this approach depends on choosing a suitable value for S. These detection mechanisms can be easily bypassed if a relatively small bias is introduced in dishonest recommendations.

### III. PROPOSED APPROACH

The main goal of Mobile Ad-hoc Network is to establish trusted connection amongst each other. We can define scenarios in MANET where a newly joined node wants to establish a secure connection with a particular CH from the set on CHs. To evaluate the trustworthiness of these CHs we have adopted indirect trust mechanism approach [19] and we have evaluated the performance and effectiveness of this technique for evaluating the trustworthiness of particular service provider (CH). The trust in short is computed on the basis of recommendations from the associated members which are frequently interacting with the service providers and the other members in the pervasive environment. In this approach, we define a scenario consisting of cluster formed using heterogeneous mobile nodes with static mobility model, in this cluster we take (for ex. 4) cluster heads, each of these CHs will perform their tasks in round robin fashion with fixed time slice i.e. at a time only one CH will be active. Now we divide our scenario into 3 phases:

A. *Interaction Phase*
B. *Request Phase*
C. *Trust Evaluation Phase*

A. *Interaction Phase*

In this phase we generate the interactions between the each cluster head and the member nodes and depending on the number of successful interactions and by considering some other communication parameters the member nodes will generate feedback values in the range from 1 to 10. These feedbacks or recommendations will be stored with the member nodes for each of the cluster head. The interaction phase can be visualized from fig 1(b).

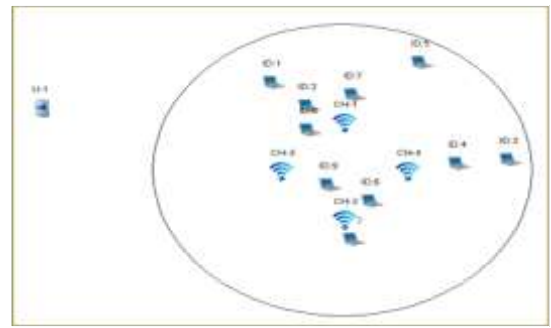

Fig. 1  MANET with static mobility model

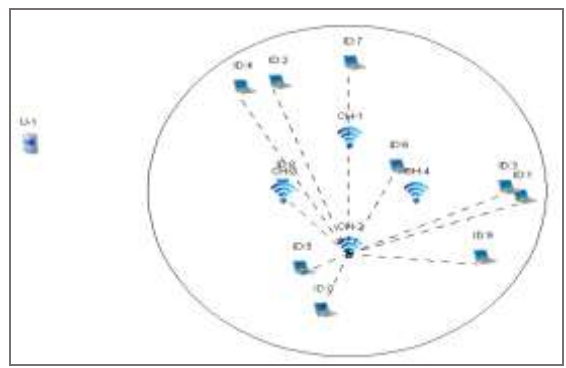

Fig. 2 Nodes interacting with CH-2 and generating feedbacks.

B. *Request Phase*

Now let us assume that some user or node wants to access a secure connection with any of the cluster head which also acts as service provider for particular service such as gateway service, ftp server, etc. but the node is not sure about the trustworthiness of the cluster heads. So when the node will enter the cluster it will request for trust ratings to all the cluster heads which then will serve this request during their respective active periods, fig 1(c).

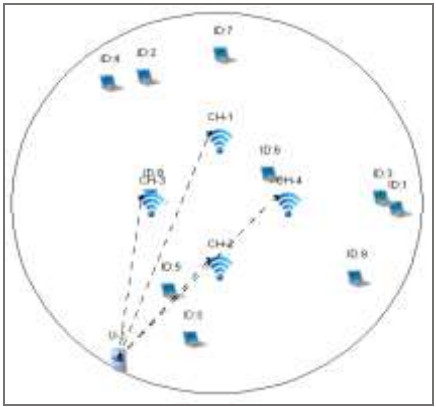

Fig. 3 Newly Entered Node requesting for Trust Value to all CHs

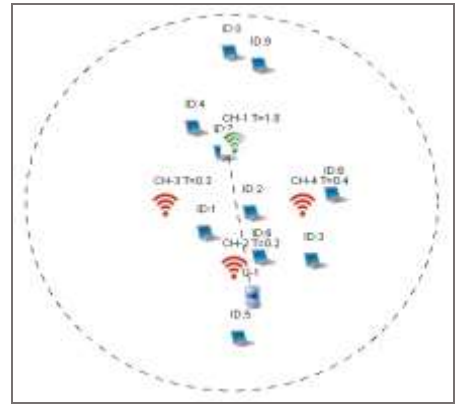

Fig. 6 Node U-1 establishes secure connection with CH having highest trust value.

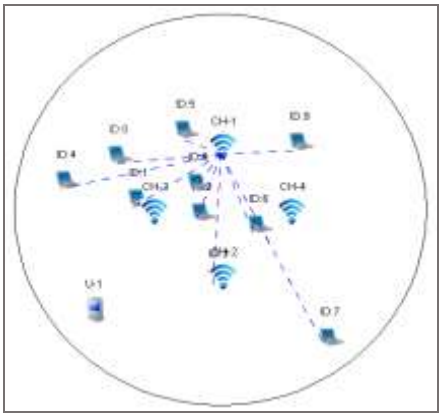

Fig. 4 CH-1 gathering recommendations from member nodes.

## C. Trust Evaluation Phase

After requesting the trust rating by the newly entered node, all the CHs will aggregate the recommendations from the member nodes in their respective active periods and will apply the indirect trust mechanism to evaluate the trust rating. The cluster head whose trust rating is greater than 0.5 will be treated as trusted but, if there are more than one trusted CHs then the CH with the highest trust rating will be chosen for establishing the secure connection. The node establishing secure connection with CH with highest trust rating can be seen in fig 1(f).

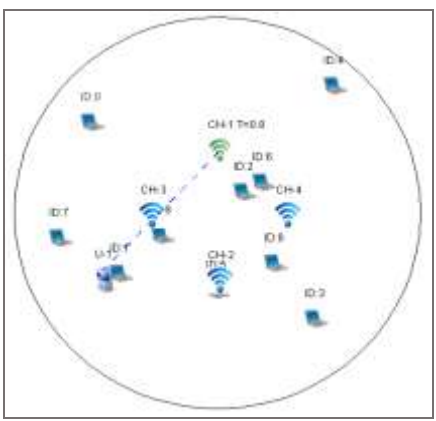

Fig. 5 Each CH reporting Trust Value to new node U-1

### IV. WORKING OF INDIRECT TRUST MECHANISM

The objective of indirect trust computation [19] is to determine the trustworthiness of an unfamiliar service provider from the set of recommendations that narrow the gap between the derived recommendation and the actual trustworthiness of the target service. In our approach, a dishonest recommendation is defined as an outlier that appears to be inconsistent with other recommendations and has a low probability that it originated from the same statistical distribution as the other recommendation in the data set. The importance of detecting outliers in data has been recognized in the fields of database and data mining for a long time. The outlier deviation-based approach was first proposed in [21], in which an exact exception problem was discussed. In [20], the author presented a new method for deviation-based outlier detection in a large database. The algorithm locates the outlier by a dynamic programming method. The approach (Algorithm 1) is based on the fact that if a recommendation is far from the median value of a given recommendation set and has a lower frequency of occurrence, it is filtered out as a dishonest recommendation. Suppose that an entity X requests to access service A. If service A has no previous interaction history with X, it will broadcast the request for recommendations, with respect to X. Let R denote the set of recommendations collected from recommenders.

$$R = \{r_1, r_2, r_3 \ldots \ldots \ldots \ldots r_n\}$$

Where 'n' is the total number of recommendations. Since smart attackers can give recommendations with little bias to go undetected, we divide the range of possible recommendation values into b intervals (or bins). These bins define which recommendations we consider to be similar to each other such that all recommendations that lie in the same bin are considered alike. 'b' has an impact on the detection rate. If the bins are too wide, honest recommendations might get filtered out as dishonest. On the other hand, if the bins are too narrow, some dishonest recommendations may appear to be honest and vice versa. For this approach authors have tuned b = 10 such that $Rc_1$ comprises all recommendations that lie between interval [0 0.1], Rc2 comprises all recommendations between interval [0.1 0.2], and so on for ($Rc_3, \ldots, Rc_{10}$). After grouping the recommendations in their respective bins, we compute a histogram that shows count fi of the recommendations falling in each bin. Let H be a histogram of a set of recommendation classes where

H(R)={<$Rc_1,f_1$>,<$Rc_2,f_2$>,<$Rc_3,f_3$>,<$Rc_4,f_4$>,<$Rc_5,f_5$>,<$Rc_6,f_6$>,<$Rc_7,f_7$>,<$Rc_8,f_8$>,<$Rc_8,f_8$>,<$Rc_9,f_9$>,<$Rc_{10},f_{10}$>}

Where fi is the total number of recommendations falling in Rci. From this histogram H(R), we remove all the recommendation classes with zero frequencies and get the domain set (Rdomain) and frequency set (f)

Rdomain= {$Rc_1, Rc_2, Rc_3, ………….. Rc_{10},$}
f = {$f_1, f_2, f_3, ………, f_{10}$}

```
Algorithm 1 Recommendation
Require: Set of Recommendations
Ensure: Rdomain_dishonest
 1: for i = 1 → 10 do
 2:    Rc_i = i/10
 3:    f_i = number of recommendations in interval [i/10 − 0.1, i/10]
 4: end for
 5: for i = 1 → 10 do
 6:    if f_i <> 0 then
 7:       Rdomain[k] = Rc_i
 8:       H[k ++] = {Rc_i, f_i}
 9:    end if
10: end for
11: x̄ = Median(Rdomain)
12: for each k in Rdomain do
13:    DF[k] = |Rdomain[k] − x̄|² / f_k   //calc deviation
14: end for
15: SRdomain = SortDesc(Rdomain, DF)
16: D_0 = ∅
17: for j = 1 to size of (SRdomain) - 1 do
18:    D_j ⋃ (SRdomain_j)
19:    SF_k = SmoothingFactor(D_j)
20: end for
21: SF_max = max (SF(D_k))
22: f_min = min freq of k in SRdomain with SF = SF_max
23: Rdomain_dishonest = all k in SRdomain with SF_k = SF_max and f_k = f_min
24: return Rdomain_dishonest
```

**Definition 1.** The dissimilarity function $DF(x_i)$ is defined as,

$$DF(x_i) = \frac{|x_i - median(x)|^2}{f_i} \quad (1)$$

Where $x_i$ is a recommendation class from a recommendation set x.

Under this mechanism, the dissimilarity value of xi is dependent on the square of absolute deviation from the median, i.e., $|x_i - median(x)|^2$. The median is used to detect deviation because it is resistant to outliers. The presence of outliers does not change the value of the median. In Equation 1, the square of absolute deviation from the median is taken to signify the impact of extremes, i.e., the farther the recommendation value $x_i$ is from the median, the larger the squared deviation is. Moreover, the dissimilarity value of $x_i$ is inversely proportional to its frequency. In Equation 1, $|x_i - median(x)|^2$ is divided by frequency fi. In this way, if a recommendation is very far from the rest of the recommendations and its frequency of occurrence is also low, Equation 1 will return a high value. Similarly, if a recommendation is close to the rest of the recommendations (i.e., similar to each other) and its frequency of occurrence is also high, Equation 1 will return a low value.

For each Rci, a dissimilarity value is computed using Equation 1 to represent its dissimilarity from the rest of the recommendations with regard to their frequency of occurrence. All the recommendation classes in Rdomain are then sorted with respect to their dissimilarity value $DF(Rc_i)$ in descending order. The recommendation class at the top of the sorted Rdomain with respect to its $DF(x_j)$ is considered to be the most suspicious one to be filtered out as dishonest recommendation. Once the Rdomain is sorted, the next step is to determine the set of dishonest recommendation classes from Rdomain set. To help find the set of dishonest recommendation classes from the set of recommendations in Rdomain, Arning et al. [21] defined a measure called smoothing factor (SF).

**Definition 2.** A SF for each SRdomain is computed as,

SF($SRdomain_j$)=C(Rdomain -$SRdomain_j$) x (DF(Rdomain)- DF($SRdomain_j$))    (2)

Where j = 1, 2, 3 . . . , m, and m is the total number of distinct elements in SRdomain. C is the cardinality function and is taken as the frequency of elements in a set {Rdomain − $SRdomain_j$}. The SF indicates how much the dissimilarity can be reduced by removing a suspicious set of recommendation (SRdomain) from the Rdomain.

**Definition 3**. The dishonest recommendation domain ($Rdomain_{dishonest}$) is a subset of Rdomain that contributes most to the dissimilarity of Rdomain and with the least number of recommendations, i.e., $Rdomain_{dishonest}$ ⊆ Rdomain. We say that SRdomainx is a set of dishonest recommendation classes with respect to SRdomain, C, and DF ($SRdomain_j$) if

SF($SRdomain_x$) >= SF($SRdomain_j$)  x, j ∈ m

In order to find out the set of dishonest recommendation Rdomaindishonest from Rdomain, the mechanism defined by the proposed approach is as follows:
•Let $Rc_k$ be the k$^{th}$ recommendation class of Rdomain and SRdomain be the set of suspicious recommendation classes from Rdomain, i.e., SRdomain ⊆ Rdomain.
•Initially, SRdomain is an empty set, $SRdomain_0$ = {}.
•Compute SF ($SRdomain_k$) for each SRdomaink formed by taking the union of $SRdomain_{k-1}$ and $Rc_k$.

$SRdomain_k = SRdomain_{k-1}$ U $Rc_k$    (3)

where k = 1, 2, 3 . . . , m − 1, and m is the distinct recommendation class value number in sorted Rdomain.

• The subset $SRdomain_k$ with the largest SF ($SRdomain_k$) is considered as a set containing dishonest recommendation classes.

• If two or more subsets in $SRdomain_k$ have the largest SF, the one with minimum frequency is detected as the set containing dishonest recommendation classes.

After detecting the set $Rdomain_{dishonest}$, we remove all recommendations that fall under the dishonest recommendation classes.

**An Illustrative Example**

To illustrate how this deviation detection mechanism filters out unfair recommendations, this section provides an example that goes through each step of our proposed approach. Let X be a service requester who has no prior experience with service provider or CH. In order to determine the trustworthiness of CH, X will get registered with CH and will request for its trust value, CH in turn will request recommendations from its peer services who have previous interaction with X. Let $R = \{r_1, r_2, r_3, \ldots, r_n\}$ be a set of recommendations received by n = 10 recommenders for service requester R. After receiving the recommendations, they are grouped in their respective bins. Table 1 shows how the received recommendations are grouped in their respective classes. After arranging the recommendations in their respective recommendation class $Rc_i$, we remove the recommendation classes with zero frequencies and calculate DF ($Rc_i$) for each recommendation class using Equation 1. Table 2 shows the sorted list of recommendation classes with respect to their dissimilarity value. In Table 2 the recommendation class $Rc_6$ has the highest deviation value, so it is taken as a suspicious recommendation class and is added to the suspicious recommendation domain (SRdomain), and its SF is calculated. Next we take the union of the suspicious recommendation domain $SRdomain_1$ and the next recommendation class in the sorted list, i.e., $Rc_4$ and calculate its SF using Equation 2. This process is repeated for each $Rc_i$ of Rdomain until $SRdomain = Rdomain − Rc_m$, where m = 6. Table 3 shows that the SF of $SRdomain_2$ has the highest value. Therefore, the recommendation classes {1.0, 0.8} in $SRdomain_3$ are considered as dishonest recommendation classes, and these recommendation classes are removed from the Rdomain.

TABLE I
FREQUENCY DISTRIBUTION OF RECOMMENDATIONS

| $Rc_i$ | Recommendation value ($rc_i$) | Frequency $f_i$ |
|---|---|---|
| $Rc_1$ | 0.1 | 2 |
| $Rc_2$ | 0.2 | 1 |
| $Rc_3$ | 0.3 | 0 |
| $Rc_4$ | 0.4 | 3 |
| $Rc_5$ | 0.5 | 0 |
| $Rc_6$ | 0.6 | 2 |
| $Rc_7$ | 0.7 | 0 |
| $Rc_8$ | 0.8 | 1 |
| $Rc_9$ | 0.9 | 0 |
| $Rc_{10}$ | 1.0 | 1 |

TABLE II
RECOMMENDATION CLASSES SORTED WITH RESPECT TO THEIR DF

| $Rc_i$ | Recommendation value $rc_i$ | Frequency $f_i$ | DF ($Rc_i$) |
|---|---|---|---|
| $Rc_6$ | 1.0 | 1 | 0.81 |
| $Rc_5$ | 0.8 | 1 | 0.49 |
| $Rc_4$ | 0.6 | 2 | 0.125 |
| $Rc_3$ | 0.4 | 3 | 0.03 |
| $Rc_2$ | 0.2 | 1 | 0.01 |
| $Rc_1$ | 0.1 | 2 | 0 |

TABLE III
SMOOTHING FACTOR COMPUTATION

| SRdomain | Rdomain-SRdomain | DF(Rdomain-SRdomain) | SF |
|---|---|---|---|
| {1.0} | {0.8,0.6,0.4,0.2,0.1} | 0.81 | 7.29 |
| {1.0,0.8} | {0.6,0.4,0.2,0.1} | 1.3 | 10.4 |
| {1.0,0.8,0.6} | {0.4,0.2,0.1} | 1.425 | 8.55 |
| {1.0,0.8,0.6,0.4} | {0.2,0.1} | 1.455 | 4.365 |
| {1.0,0.8,0.6,0.4,0.2} | {0.1} | 1.465 | 2.93 |

V. PERFORMANCE EVALUATION

In this section, we evaluate our model in a simulated cluster based MANET environment. We carry out different sets of experiments to demonstrate the effectiveness of the proposed model against different attack scenarios (BM attack, BS attack, and RO attack). Results indicate that the model is able to respond to all three types of attack when the percentage of malicious recommenders is varied from 10% to 40%. We have also studied the performance of the model by varying the offset introduced by the malicious recommender in their recommended trust value. It was observed that the performance of the models decreases only when the percentage of malicious recommenders is above 30% and the mean offset between the honest and dishonest recommendation is minimum (0.2).

**Experimental setup**

We simulate a MANET environment using a Java based simulator, where nodes (offering and requesting services) are continuously joining and leaving the environment. The nodes are categorized into two groups, i.e., agents offering services as service provider nodes (SPN) and nodes consuming services as service requesting nodes (SRN). We conduct a series of experiments for a new SRN to evaluate the trustworthiness of an unknown SPN by requesting recommendation from other SPNs in the environment. All SPNs can also act as recommending agents (RA) for other SPNs. The RA gives recommendations, in a continuous range [0 1], for a given SPN on the request of a SRN. The RA can either be honest or dishonest depending on the trustworthiness of its recommendation. An honest RA truthfully provides recommendation based on its personal experience, whereas a dishonest RA insinuates a true experience to a high, low, or erratic recommendation with a malicious intent. The environment is initialized with set numbers of honest and dishonest recommenders (N = 10 to 100).

### A. *Experiment 1: validation against attacks*

To analyse the effectiveness of the proposed approach, three inherent attack scenarios (bad mouthing, ballot stuffing, and random opinion attack) for recommendation models have been implemented in the above defined simulation environment.

#### *1) Bad mouthing attack*

BM is one in which the intention of the attacker is to send malicious recommendations that will cause the evaluated trustworthiness of an entity to decrease. Let us suppose that the service provider asks for recommendations regarding an unknown service provider node CH-1. In this experiment we assume that a certain percentage of the recommenders are dishonest and launch a BM attack against (CH-1) by giving dishonest recommendations. It is assumed that the actual trust value of CH-1 is 0.9. At the initial step of the simulation, the environment has 10% dishonest RA who attempt to launch a bad mouthing attack against A by providing low recommended trust values (between the range [0 0.3]). To elaborate the efficacy of the proposed approach, we vary the percentage of dishonest recommenders from 10% to 40%. Figure 7 a, b, c, d shows the SF calculated for each SRdomain. It is shown that in each case the proposed approach is able to detect the set of bad mouthers giving low recommendation between 0.1 and 0.3, also the experiment has been performed for N=30. For example, in Figure 7a when the percentage of dishonest recommenders is 10%, the SRdomains and respective SF values are as follows:

| | |
|---|---|
| SRdomain 1{0.1} | 23.49 |
| SRdomain 2{0.1,0.2} | 40.6 |
| SRdomain 3{0.1,0.2,0.3} | 52.38 |
| SRdomain 4{0.1,0.2,0.3,0.8} | 34.9992 |
| SRdomain 5{0.1,0.2,0.3,0.8,0.9} | 13.6171 |

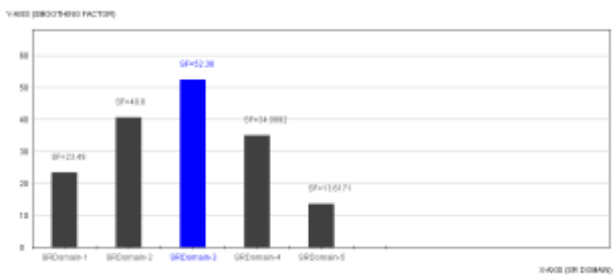

(a)

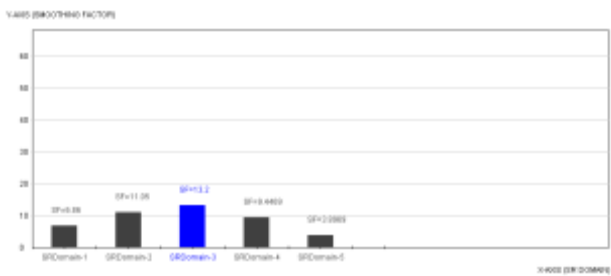

(b)

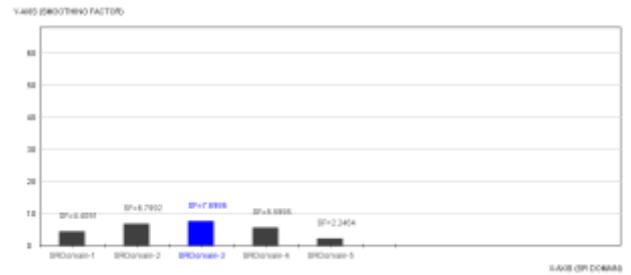

(c)

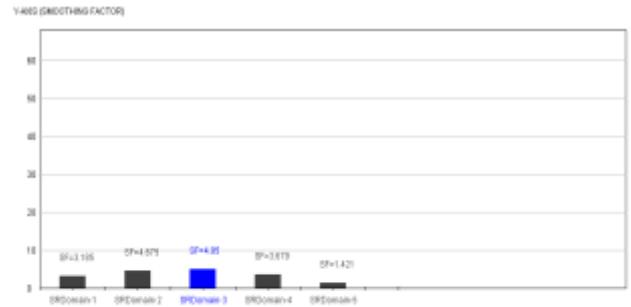

(d)

Fig. 7 Detecting Attacks. (a) BM, 10% dishonest recommender. (b) BM, 20% dishonest recommender. (c) BM, 30% dishonest recommender. (d) BM, 40% dishonest recommender.

Since the SF of SRdomain 3 has the highest value, the recommendation classes {0.1, 0.2, 0.3} are considered as dishonest recommendation classes, and the recommendations that belong to these recommendation classes are considered as dishonest recommendations.

#### *2) Ballot stuffing attack*

BS is one in which the intention of the attacker is to send malicious recommendations that will cause the evaluated trustworthiness of an entity to increase. Let us suppose that the service requester asks for recommendations regarding an unknown service provider CH-1. It is assumed that the actual trust value of CH-1 is 0.3. A certain percentage of recommenders providing the recommendation to the service provider are dishonest and gives a high recommendation value between 0.8 and 1.0, thus launching a BS attack. We evaluate the proposed approach by varying the percentage of dishonest recommenders from 10% to 40%. Figure 7 e, f, g, h shows the SF values for SRdomains in each case. It is evident from the results that the model is able to detect dishonest recommendations even when the percentage of dishonest recommendations is 40%. From Figure 7h (when the percentage of dishonest recommendations is 40%), the SF values of each SRdomain are as follows:

| | |
|---|---|
| SRdomain 1{1.0} | 5.265 |
| SRdomain 2{1.0,0.8} | 8.413 |
| SRdomain 3{1.0,0.8,0.9} | 8.884 |
| SRdomain 4{1.0,0.8,0.9,0.3} | 6.006 |
| SRdomain5{1.0,0.8,0.9,0.3,0.2} | 3.013 |

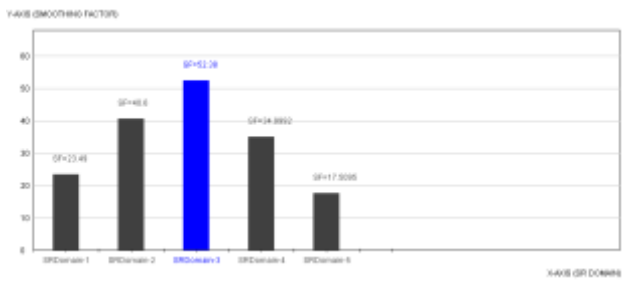

(e)

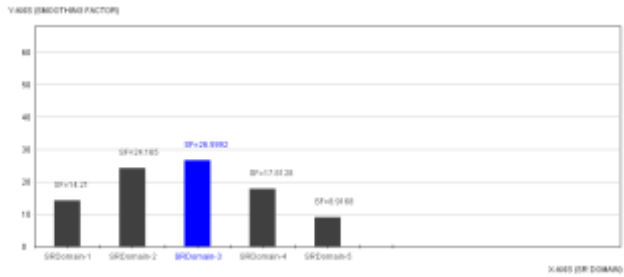

(f)

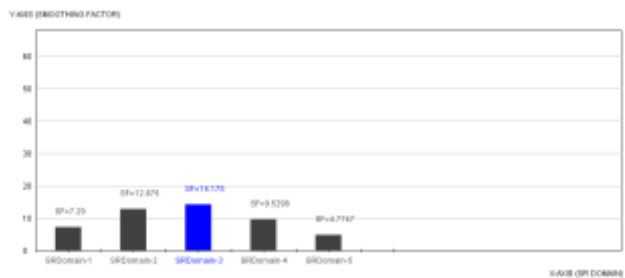

(g)

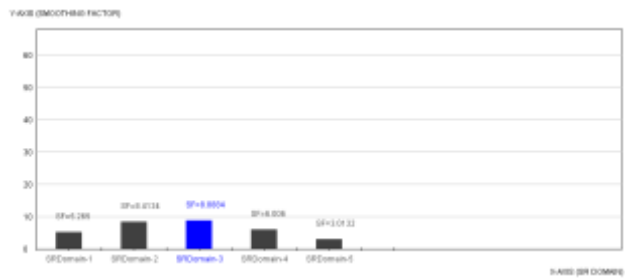

(h)

Fig. 7 Detecting Attacks. (e) BS, 10% dishonest recommender. (f) BS, 20% dishonest recommender. (g) BS, 30% dishonest recommender. (h) BS, 40% dishonest recommender.

The proposed approach is able to detect the dishonest recommendations as SRdomain3 with the highest SF value of 8.88.

*3) Random opinion attack*

RO attack is one in which the malicious recommender gives the recommendations randomly opposite the true behavior of the entity in question. Let us suppose that the recommenders launch a RO attack while providing recommendations for a service provider node CH-1. The dishonest recommenders provide either very low recommendations (0.1 to 0.2) or very high recommendations (0.8 to 1.0). We vary the percentage of dishonest recommenders from 10% to 40% for the experiment. The SF values for the respective SRdomains in each case are shown in Figure 7 i, j, k, l. The proposed approach successfully detects random opinion attack and is able to filter out the dishonest set of recommenders in each case. From Figure 7i (when the percentage of dishonest recommendations is 10%), the SF values of each SRdomain are as follows:

| | |
|---|---|
| SRdomain 1{0.1} | 7.25 |
| SRdomain 2{0.1,1.0} | 11.48 |
| SRdomain 3{0.1,1.0,0.2} | 15.39 |
| SRdomain 4{0.1,1.0,0.2,0.3} | 9.251 |
| SRdomain 5{0.1,1.0,0.2,0.3,0.4} | 6.448 |
| SRdomain 6{0.1,1.0,0.2,0.3,0.4,0.7} | 2.939 |

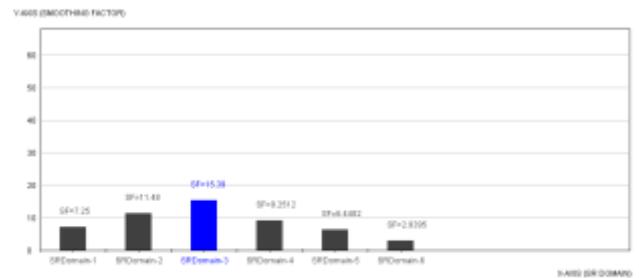

(i)

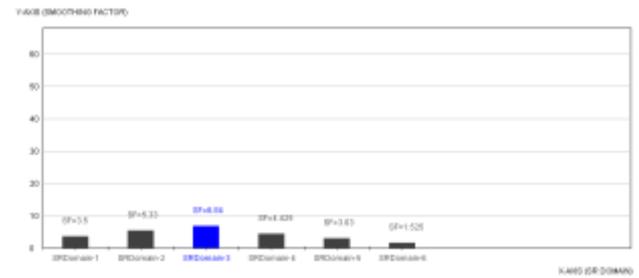

(j)

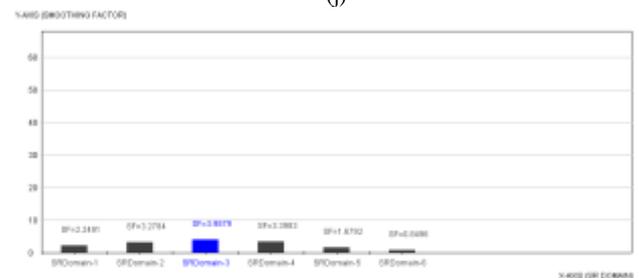

(k)

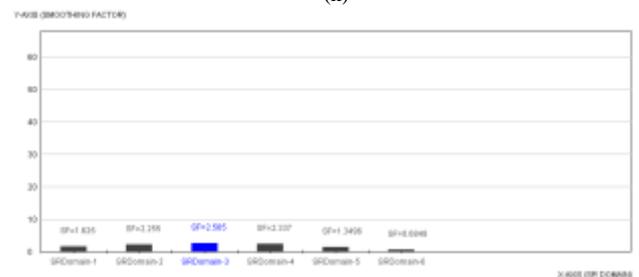

(l)

Fig. 7 Detecting Attacks. (**i**) RO, 10% dishonest recommender. (**j**) RO, 20% dishonest recommender. (**k**) RO, 30% dishonest recommender. (**l**) RO, 40% dishonest recommender.

*B. Experiment 2: validation against deviation*

The detection rate of unfair recommendations by varying the number of malicious recommenders cannot fully describe the performance of the model as the damage caused by different malicious recommenders can be very different depending on the disparity between the true recommendation and unfair recommendation (i.e., offset). The offset introduced by the attackers in the recommended trust value is a key factor in instilling deviation in the evaluated trust value of SPN. We have carried out a set of experiments to observe the impact of different offset values introduced by different malicious recommenders on the final trust value. We define mean offset (MO) as the difference between the mean of honest recommendations and the mean of dishonest recommendations. For the experiment, we have divided MO into four different levels L1 = 0.1, L2 = 0.2, L3 = 0.4, and L4 = 0.8. It is assumed that the actual trust value of SPN is less than equal to 0.5, and the dishonest recommender's goal is to boost the recommended trust value of SPN (BS attack). The experiment was conducted in four different rounds by varying the MO level from L1 to L4 (i.e., from maximum to minimum). In each round, the recommended trust value is computed with different percentages of dishonest recommenders (10%, 20%, 30%, and 40%).

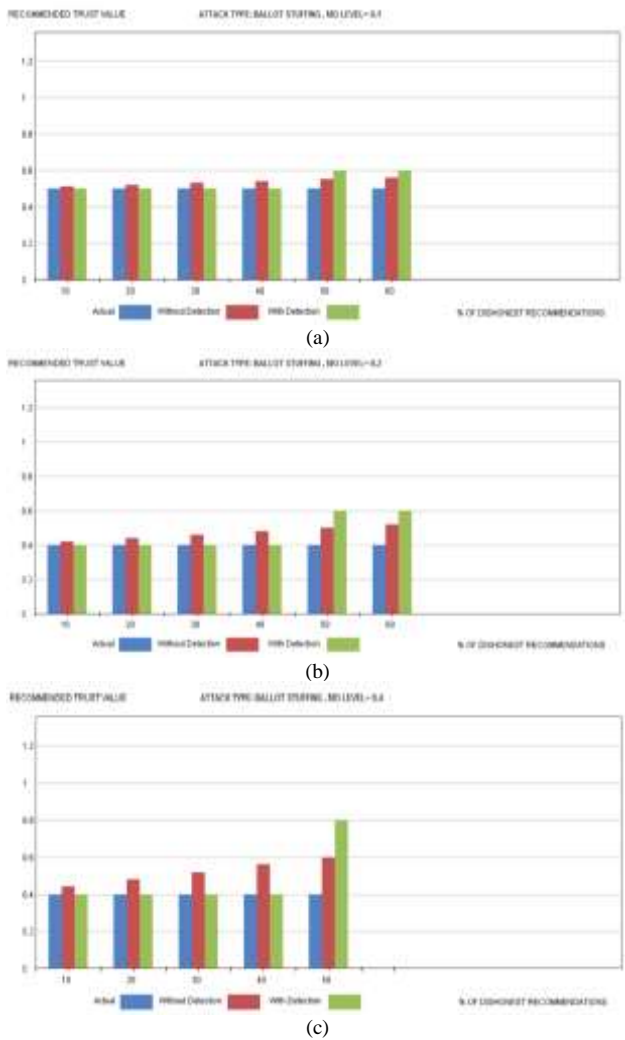

(a)

(b)

(c)

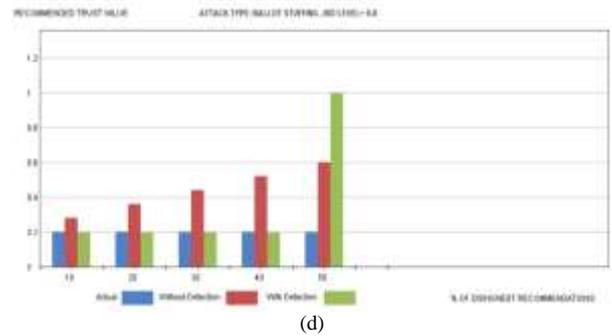

(d)

Fig. 8. Mean Offset Analysis. (a) Mean Offset L1. (b) Mean Offset L2. (c) Mean Offset L3. (d) Mean Offset L4

### C. Comparison with existing approaches

To illustrate the effectiveness of the proposed deviation-based approach in detecting dishonest recommendations, we have compared our approach with other approaches proposed in the literature based on quartile [15], control limit chart [17], and iterative filtering [18] to detect dishonest recommendations in indirect trust computation. A set of experiments has been carried out by applying the approaches to detect dishonest recommendations in two different scenarios. For the first set of experiments, we assume that a certain percentage of the recommenders are dishonest and launch bad mouthing attack by giving recommendations between 0.1 to 0.3. For the second set of experiments, the dishonest recommenders are assumed to give a high recommendation value in range [0.8, 1.0], thus launching a ballot stuffing attack. In both set of experiments, the percentage of dishonest recommenders is varied from 10% to 45%. For comparison, we have used Matthews's correlation coefficient (MCC) to measure the accuracy of all four approaches in detecting dishonest recommendations [22]. MCC is defined as a measure of the quality of binary (two-class) classifications. It takes into account true and false positives and negatives. The formula used for MCC calculation is

$$MCC = \frac{(TP \times TN) - (FP \times FN)}{\sqrt{(TP+FP)(TP+FN)(TN+FP)(TN+FN)}}$$

Where TP is the number of true positives, TN is the number of true negatives, FP is the number of false positives, and FN is the number of false negatives. MCC returns a value between − 1 and 1 (1 means perfect filtering, 0 indicates no better than random filtering, and − 1 represents total inverse filtering). To avoid infinite results while calculating MCC, it is assumed that if any of the four sums (TP, FP, TN, and FN) in the denominator is zero, the denominator is arbitrarily set to one. The Figure 4 shows the comparison of MCC values of the proposed approach with different model with varying percentage of dishonest recommendations (from 10% to 40%). According to the results, the proposed approach can effectively detect dishonest recommendations evident from a constant MCC of +1 for both sets of experiments. On the other hand, in [17], in the case of bad mouthing attack (Figure 9a), MCC increases slowly as the percentage of dishonest recommenders increases from 10% to 30% but then decreases promptly to 0 as the percentage of dishonest recommender increases from 30% to 45%. The same behavior was observed in the case of ballot stuffing attack (Figure 9b). In [18], when the percentage of dishonest recommender increases to 40%, the MCC rate starts to decrease as well. Thus, all three approaches ([17, 18], and [13]) fail to achieve perfect filtering of dishonest recommendation as the percentage of dishonest recommenders increases.

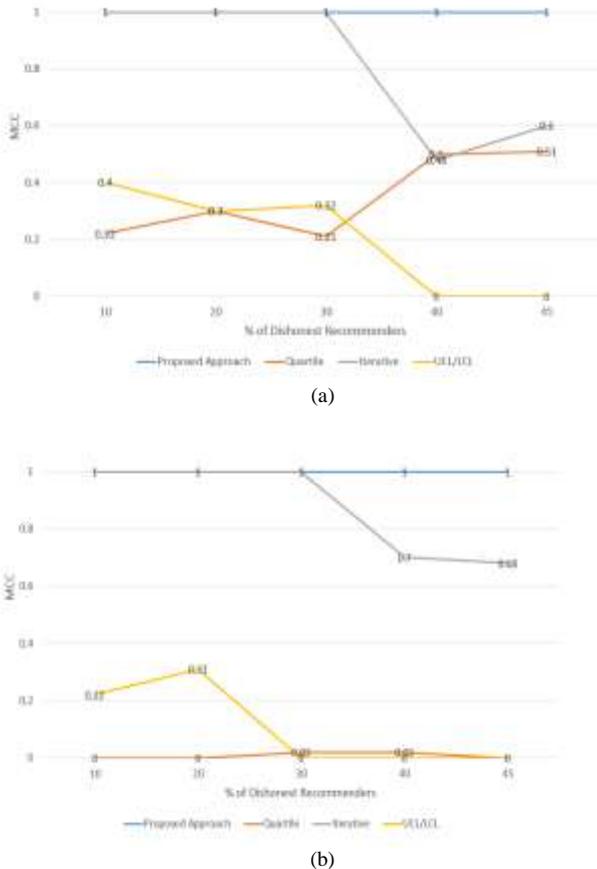

(a)

(b)

Fig. 9. Filtering accuracy in terms of MCC. (a) Bad mouthing Attack. (b) Ballot Stuffing Attack [19].

For an in-depth analysis of [17, 18], and [13], false positive rate (FPR) and false negative rate (FNR) are computed for using the following equations:

$$FPR = \frac{FP}{FP + TN}$$

$$FNR = \frac{FN}{FN + TP}$$

The value of FPR and FNR lies between [0 1]. The lower value of FPR and FNR indicates better performance. Figure 10a shows the comparison of FNR and FPR of [17] with the proposed approach based on the results accumulated after the experiments for BM attack. Although the FPR of [17] remains consistent at zero when the percentage of dishonest recommendations is the value of FPR and FNR lies between [0 1]. The lower value of FPR and FNR indicates better performance. Figure 10a shows the comparison of FNR and FPR of [17] with the proposed approach based on the results accumulated after the experiments for BM attack. Although the FPR of [17] remains consistent at zero when the percentage of dishonest recommendations is increased from 10% to 40%, at the same time, its FNR progressively increases and reaches its maximum value at 40%. Similarly, Figure 10b shows that [18] maintains zero FPR and FNR until dishonest recommenders are less than 30% of the total recommenders. However, as the number of dishonest recommenders increases above 30%, the model behaves poorly by showing a rapid increase in FPR and FNR. Figure 10c shows that although the FPR of [13] improves as the percentage of dishonest recommenders' increases, simultaneously, the FNR starts to grow rapidly for percentages greater than 20%. On the contrary, the proposed approach maintains zero FNR and FPR even when the percentage of dishonest recommenders reaches 40%. Figure 11a explicates the results observed from the performance of [17] under ballot stuffing attack. The approach maintains zero FPR throughout the experiment; however, it filtered out a high number of honest recommenders as dishonest, evident from the high FNR. Similarly, the performance of [18] remains stable until the percentage of dishonest recommenders remains below 30% (Figure 11b). However, the approach also shows a rapid growth in FNR as the percentage of dishonest recommenders increases above 30%. Figure 11c shows that [13] is completely unable to detect ballot stuffing. The approach shows a high FPR even at low percentages of dishonest recommenders. It can be seen from the results of Figures 5 and 6 that the proposed approach remains resistant to the attack under both experiments (as the FPR and FNR remains zero), thus outperforming other approaches.

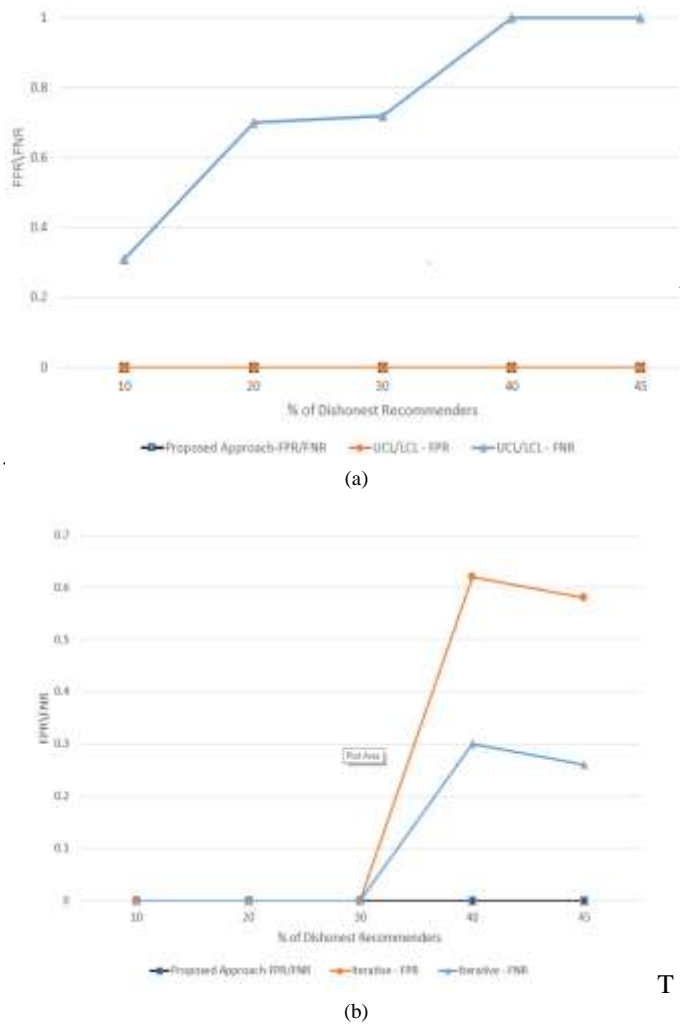

(a)

(b)

T

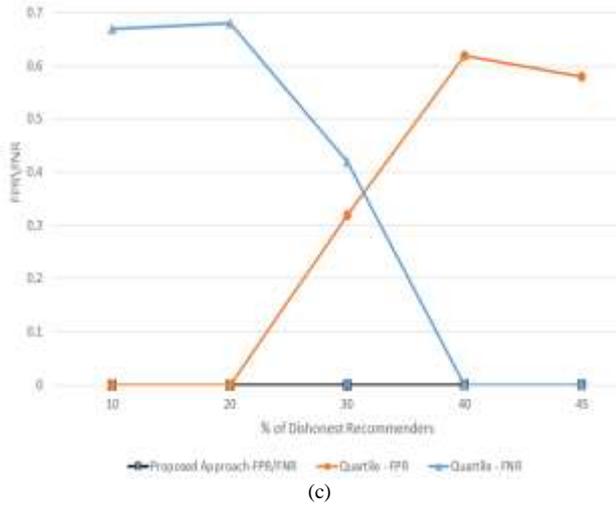

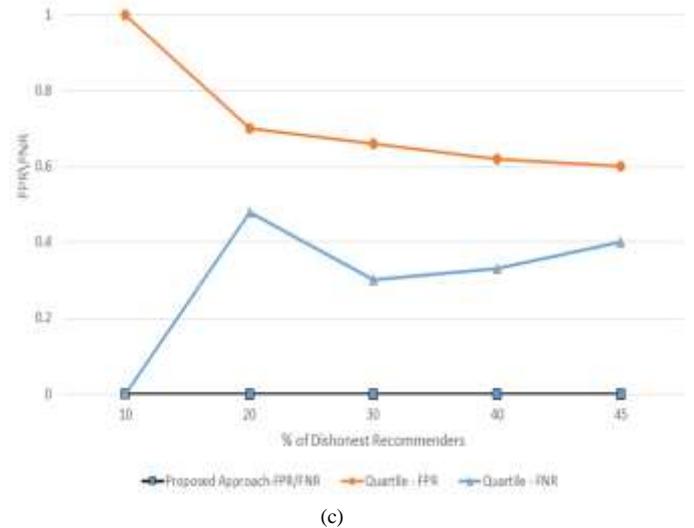

Fig. 10. FPR vs FNR (in Bad Mouthing Attack). (a) Control Limit Chart. (b) Iterative Filtering. (c) Quartile Filtering [19]

Fig. 11. FPR vs FNR (in Ballot Stuffing Attack). (a) Control Limit Chart. (b) Iterative Filtering. (c) Quartile Filtering. [19]

From the above discussion, we can conclude that both [17] and [13] perform poorly in the presence of increasing percentage of dishonest recommenders. It is also observed that [18] performs well provided that the recommendation threshold is selected appropriately. On the contrary, the proposed approach is not reliant on any external parameter and is able to detect 100% dishonest recommenders provided that they are in the minority (<50%).

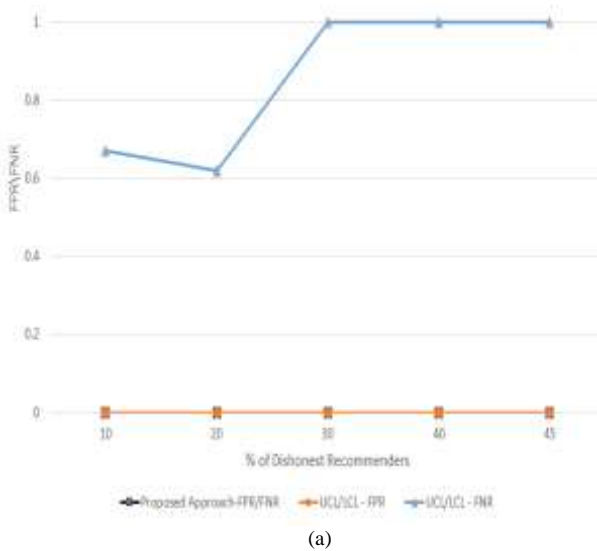

(a)

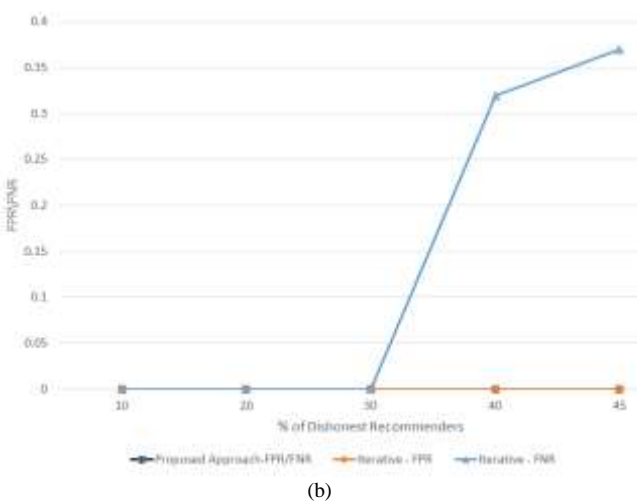

(b)

## VII. CONCLUSIONS

An application of indirect trust mechanism in MANET environment is proposed in the above work. The main focus in this present work was to detect dishonest recommendations based on their dissimilarity value from the complete recommendation set. Since median is resistant to outlier, we have proposed a dissimilarity function that captures how dissimilar a recommendation class is from the median of the recommendation set. The algorithm uses a smoothing factor which detects malicious recommendations by evaluating the impact on the dissimilarity metric by removing a subset of recommendation classes from the set of recommendations. Experimental evaluation shows the effectiveness of our proposed method in filtering dishonest recommendations in comparison with the base model. Results show that the proposed method is successfully able to detect dishonest recommendations by utilizing absolute deviation from the median as compared to the base technique which tends to fail as the percentage of dishonest recommendations increases. We have carried out a detailed comparative analysis with the base approach by varying the percentage and the offset introduced by the dishonest recommendations. Results that indicate improved performance of the proposed approach, which is able to produce 70% detection rate at a minimum offset of 0.2, have been shown. On the contrary, the base approach is unable to detect any dishonest recommendations at all. It is also shown that for different attacks (bad mouthing, ballot stuffing, and random opinion attack), the proposed method successfully filters out dishonest recommendations. A comparison between existing approaches and the proposed approach is also presented, which clearly shows the better performance of the proposed approach. In our future work, we will try to incorporate fuzzy or rule based engine to evaluate the trust value from the recommendation sets.